\documentclass[preprint]{imsart}

\usepackage{amsthm,amsmath,natbib,array,amssymb,
bm,graphicx,booktabs,multirow}

\DeclareMathOperator*{\argmax}{arg\!max}
\DeclareSymbolFont{bbold}{U}{bbold}{m}{n}
\DeclareSymbolFontAlphabet{\mathbbold}{bbold}
\RequirePackage[colorlinks,citecolor=blue,urlcolor=blue]{hyperref}

\usepackage{pdflscape}

\begin{document}

\begin{frontmatter} 

\title{Sequential Advantage Selection for Optimal Treatment Regimes}
\runtitle{SAS for Optimal Treatment Regimes}


\begin{aug}
\author{\fnms{Ailin} \snm{Fan}\ead[label=e1]{afan@ncsu.edu}},
\author{\fnms{Wenbin} \snm{Lu}\ead[label=e2]{lu@stat.ncsu.edu}}
\and
\author{\fnms{Rui} \snm{Song}\ead[label=e3]{rsong@ncsu.edu}}
\affiliation{North Carolina State University}


\runauthor{A. Fan et al.}
\end{aug}

\begin{abstract}
Variable selection for optimal treatment regime in a clinical trial or an observational study is getting more attention. Most existing variable selection techniques focused on selecting variables that are important for prediction, therefore some variables that are poor in prediction but are critical for decision-making may be ignored. A qualitative interaction of a variable with treatment arises when treatment effect changes direction as the value of this variable varies. The qualitative interaction indicates the importance of this variable for decision-making. \citet{gunter2011variable} proposed S-score which characterizes the magnitude of qualitative interaction of each variable with treatment individually.  In this article, we developed a sequential advantage selection method based on the modified S-score. Our method selects qualitatively interacted variables sequentially, and hence excludes marginally important but jointly unimportant variables { or vice versa}. The optimal treatment regime based on variables selected via joint model is more comprehensive and reliable.
With the proposed stopping criteria, our method can handle a large amount of covariates even if sample size is small. Simulation results show our method performs well in practical settings. We further applied our method to data from a clinical trial for depression.
\end{abstract}


\end{frontmatter}

\section{Introduction}
Personalized medicine is emerging as a new strategy for treatment which takes individual heterogeneity in disease severities, background characteristics, clinical measurements {  and genetic information} into consideration. In this paradigm, treatment duration, dose and type are adjusted over time and tailored according to an individual's information, aiming to optimize the effectiveness of treatment. This is different from the traditional ``one-size-fits-all" treatment, which ignores the long-term benefits and individual heterogeneities. Great interest lies in finding optimal treatment regimes based on data from clinical trials and observational studies \citep[e.g.][]{murphy2003optimal,robins2004optimal,moodie2007demystifying}.

An optimal treatment regime could involve one treatment decision or a sequence of treatment decisions at multiple stages. The latter one is referred to as the optimal dynamic treatment regime, which is a sequence of decision rules tailored through time to individual's information, and will maximize the final expected response when implemented.
{A large number of works have been developed to derive} optimal dynamic treatment regimes based on data from clinical trials and observational studies. {For example,} marginal {structure} models \citep{robins1997causal,murphy2001marginal} allow estimation of the mean response {under} a dynamic treatment regime. Q-learning \citep{watkins1989,watkins1992q,murphy2005experimental,
zhao2011reinforcement,chakraborty2010inference,song2011penalized} and A-learning \citep{murphy2003optimal,robins2004optimal} are {two popular backward induction methods for deriving optimal dynamic treatment regimes: the former builds regression models for the so-called Q functions while the latter is based on modeling contrast functions.} More recently, {\cite{zhang2012robust} proposed a value-function based optimization algorithm to find the best treatment regime in a specified class of treatment decision rules based on the inverse propensity score weighted (IPSW) estimation of the value function. They further extended the method to an augmented IPSW  estimation, which enjoys the double robustness property. \cite{zhao2012estimating} proposed an outcome-weighted learning method, recasting the estimation method of \cite{zhang2012robust} in a classification framework, and developed an outcome weighted support vector machine to estimating optimal treatment regimes.}

In practice, the sequential multiple assignment randomized trial (SMART) is an experimental design useful for deriving optimal dynamic treatment regimes \citep{murphy2005experimental,qian2013dynamic}. The SMART designs have been used for some chronic or relapsing diseases. Examples include the Sequenced Treatment Alternatives to Relieve Depression (STAR*D) trial \citep{fava2003background,rush2004sequenced}, the Clinical Antipsychotic Trials of Intervention Effectiveness (CATIE) for treatment of Alzheimer's disease \citep{schneider2001national} and the psychosocial treatments study for Attention Deficit Hyperactivity Disorder (ADHD) in children and adolescents \citep{pelham1998empirically}.
As our ability to collect individual's information is growing, there are more and more covariates measured and available in clinical studies. For example, a clinical trial may collect a large amount of information on patient's demographics, medical history, intermediate outcomes and side effects. However, it may be expensive or time-consuming to collect all these information in clinical practice, and redundancy in covariates information may impair the accuracy of optimal treatment decisions as well as the interpretations. Hence a natural problem for finding optimal treatment regimes is to select important covariates that could facilitate us to make treatment decisions. 

{Our work is motivated from the Sequenced Treatment Alternatives to Relieve Depression (STAR*D) study.} The STAR*D study is a randomized multistage clinical trial of patients with nonpsychotic major depressive disorder. As there are a lot of antidepressant medications but not a single treatment is universally effective, this study aims to determine which treatment strategies, in what order or sequence, provides the optimal treatment effect. 
However, there are a large amount of covariates collected at baseline, including patient's demographic characteristics, medical history, etc. For treatments at the second level and higher, there are also lots of intermediate medical measurements that are available for making the treatment decisions at next level. It's hard to select covariates useful for making treatment decisions from such a large amount of covariates based on expert opinion only, hence variable selection for identifying the optimal treatment regime is needed.

Variable selection is an important area in modern statistical research. However, current variable selection techniques focused on selecting variables relevant for prediction, which may not be suitable for selecting variables relevant for decision making. Variables that are vital for decision making may be neglected by techniques targeting on prediction due to the small predictive abilities of the interactions of these variables with treatment. In medical decision-making setting, \citet{gunter2011variable} distinguished between {\it predictive} variables, which help to increase accuracy of estimator, and {\it prescriptive} variables, which facilitate prescription of optimal treatment regimes. A prescriptive variable must have qualitative interaction with treatments, that is, treatment effects change direction when the value of this variable varies. Our goal of this paper is to develop a variable selection method to identify prescriptive variables for implementing optimal treatment regimes, especially when the number of covariates is large.

Scarce research have been carried out to study the variable selection techniques  for decision making. 
Qualitative interaction tests \citep{gail1985testing,piantadosi1993comparison,yan2004test} have been used to test a small number of expert determined pre-specified interactions. However, many of the tests were designed for testing only qualitative interactions between categorical variables and treatments. Moreover, when the number of covariates is large these tests are too conservative when controlling the error rate for multiple testing.
Penalized methods have also been studied to identify variables important for making treatment decisions.
\citet{qian2011performance} developed a two-stage procedure in the framework of Q-learning, where they first estimated the conditional mean response using $L_1$ penalized least squares, and then derived the estimated optimal treatment regime from estimated conditional mean. The use of $L_1$ penalized least squares leads to estimated optimal treatment regimes with fewer variables needed.
\citet{lu2013variable} proposed a penalized least squares regression framework to select important variables for making treatment decisions, which corresponds to a form of A-learning. Shrinkage penalties were incorporated on the interaction terms such that variables relevant to decision making were selected.
However, both of these penalized methods do not directly target on selecting prescriptive variables, and hence may be inappropriate for selecting variables that are important for decision making.
\citet{gunter2011variable} proposed variable selection methods for qualitative interactions, where two variable-ranking quantities were presented. These quantities characterize qualitative interactions, which depend on two factors: the magnitude of interaction and the proportion of patients for whom the optimal treatment changes given the knowledge of the variable. However, these quantities may identify too many covariates as the potential prescriptive variables when there are a large number of covariates. Also these quantities examine the effect of each variable individually, which may ignore the correlation between covariates and thus lead to identification of spurious variables or miss some important prescriptive variables for decision making.

In this paper, we proposed a new sequential advantage variable selection method. Our goal is to select variables with qualitative interactions to derive the optimal treatment regime, but our method is based on {\it sequential advantage}, which is the additional improvement of the value based on the estimated optimal treatment regime with a new variable included. This sequential advantage selection method takes variables already in the model into account and judge whether to include a new variable or not by the additional information provided by this variable on decision making. 
As our method selects relevant variables sequentially, it can handle a large number of covariates even if the sample size is small.
Another merit of the sequential selection procedure is that it does not include those variables which are marginally important but jointly unimportant and hence avoids redundant information in the model for decision making.
We also proposed a stopping criteria based on {proportion of incremental sequential advantages} to decide how many variables to be included for decision making.

The paper is organized as follows. In Section 2 we introduce the framework for identifying optimal treatment regimes, and the S-score method for selecting prescriptive variables. Section 3 provides the proposed sequential advantage selection for variable selection in optimal treatment decision making. We demonstrate their performance in Section 4 by simulation studies in various scenarios, and illustrate these methods using data from the STAR*D clinical trial in Section 5.

\section{Variable Selection for Optimal Treatment Regime}
\subsection{Optimal Treatment Regime}

Consider a clinical trial or an observational study where point exposure treatments are given to $n$ subjects sampled from the population of interest, and assume that there are two possible treatment options assigned to each subject. Denote $A$ as the treatment and assume possible values of $A$ are coded as $\{0,1\}$, which is in accordance with two treatment options. We aim to find the optimal treatment regime based on $p-$dimensional covariates of a subject $\mathbf X=(X_1,X_2,...,X_p)^T$. Following a treatment $A$ given to the subject, a response $Y$ could be obtained. Suppose larger response is preferred. 
The observed data can be summarized as $({\mathbf X}_i,A_i,Y_i),~i=1,...,n$, where ${\mathbf{X}_i=(X_{i1},...,X_{ip})^T}$. 
A treatment regime $d({\mathbf X})$ is a mapping from $\mathcal{X}$, the space of covariates $\mathbf X$, to $\mathcal{A}$, the space of action $A$. Our goal is to find the treatment regime $d^{opt}({\mathbf X})$ that can maximize the expected mean response. 

In this context, we want to identify the optimal treatment regime mapping from $\mathcal{X}$ to $\{0,1\}$. To put it formally, we need to introduce potential outcomes $Y^*(1)$ and $Y^*(0)$, which are the outcomes that would be observed if a subject is assigned to treatment 1 or 0. With the concepts of potential outcomes, an optimal treatment regime can be defined as $d^{opt}(\mathbf X)=\argmax_{d\in \mathcal{D}}\mathbb E[Y^*(d(\mathbf X))]$, where $\mathcal D$ is the collection of all possible treatment regimes, i.e., all possible mappings from $\mathcal X$ to $\mathcal A$. 

Two assumptions are essential to make identifying optimal treatment regimes possible. First is the stable unit treatment value assumption \citep{rubin1978bayesian}: $Y=I(A=0)Y^*(0)+I(A=1)Y^*(1)$. That is, an individual's outcome is the same as the potential outcome for the assigned treatment, and is not influenced by other individuals' treatment allocations. Second is the no unmeasured confounders assumption: $\{Y^*(0),Y^*(1)\}\perp A|\mathbf X$, i.e., the treatment assignment is independent of the potential outcomes conditional on {observed covariate information} \citep{robins1997causal}. With these two assumptions, it is straightforward to show that $$\mathbb E[Y^*(d(\mathbf X))]=E_{\mathbf X}\left\{E(Y|\mathbf X,A=1)d(\mathbf X)+E(Y|A=0,\mathbf X)[1-d(\mathbf X)]\right\},$$ that is, $\mathbb E[Y^*(d(\mathbf X))]$ can be expressed by the observed data.
Hence the optimal treatment regime could be derived as:
$$d^{opt}(\mathbf{x})=\argmax_{a\in \mathcal{A}} \mathbb{E}(Y|\mathbf{X} =\mathbf{x},A=a).$$
That is, for each possible value of covariate $\mathbf{X}$, we'll find the treatment $a$ that maximizes the expected response. Because the distribution of $(\mathbf{X},A,Y)$ is unknown in practice, we could only estimate $\mathbb{E}(Y|\mathbf{X} =\mathbf{x},A=a)$ given a posited model based on the observed data.

\subsection{S-Score Method}

When there are a large amount of covariates, only some of the covariates need to be considered in the model $\mathbb{E}(Y|\mathbf{X} =\mathbf{x},A=a)$ as independent predictors for making optimal treatment decision. \citet{gunter2011variable} mentioned that current variable selection techniques pay more attention to selecting the predictive variables, but are likely to neglect some important prescriptive variables. 
\citet{parmigiani2002modeling} proposed value of information using the experimental outcome to help evaluating the treatment effect. 
\citet{gunter2011variable} proposed S-score based on the value of information, which is a quantity of a variable that characterizes the degree of qualitative interaction. The ranking of S-scores shows the ranking of possibilities of different covariates to be prescriptive variables. To put it formally, the S-score for a univariate covariate $X_j$, $j=1,...,p$, is defined as:
\begin{equation} \label{Sscore}
S_j=\sum^n_{i=1}\left\{\max_a[\hat{\mathbb{E}}(Y|X_j=x_{ij},A=a)]-\hat{ \mathbb{E}}(Y|X_j=x_{ij},A=\hat a)\right\}.
\end{equation}
Here $\hat{\mathbb{E}}(Y|X_j=x_{ij},A=a)$ is an estimator of $\mathbb{E}(Y|X_j=x_{ij},A=a)$, the expected response of $Y$ given covariate $X_j$ and treatment $A$, and $\hat a$ is the optimal treatment regime { without considering the effect of the covariates}, that is, $\hat a=\argmax_a \hat {\mathbb{E}}(Y|A=a)$. For example, if the model for $\mathbb{E}(Y|X_j,A)$ is $\mathbb{E}(Y|X_j,A)=\beta_0+\beta_1 X_j+\beta_2 A +\beta_3 X_j A$, then $(\beta_0,\beta_1,\beta_2,\beta_3)^T$ can be estimated by the least squares estimates, denoted as $(\hat\beta_0,\hat\beta_1,\hat\beta_2,\hat \beta_3)^T$. S-score for $X_j$ then has the following expression:
\begin{equation}\label{eq:Sscoreex}
S_j=\sum_{i=1}^n (\hat\beta_2+\hat \beta_3 x_{ij})\left[\mathbf{1}(\hat\beta_2+\hat \beta_3 x_{ij}\geq 0)-\hat a\right].
\end{equation}

We note that $S_j$ is always non-negative. The first and second terms of $S_j$ in (\ref{Sscore}) are obtained based on the same model of $Y$ given covariate $X_j$ and treatment $A$, but the first term adopts the optimal treatment based on the posited model for ${\mathbb{E}}(Y|X_j=x_{ij},A=a)$, while the second term adopts the optimal treatment based on no covariate information. Hence only the interaction term of those individuals whose optimal treatment based on $X_j$ is different from the optimal treatment $\hat a$ that based on no covariate information will give a contribution to the S-score. As stated in \citet{gunter2011variable}, the S-score captures both the magnitude of interaction and the proportion of subjects whose optimal treatment changes. For example, in equation (\ref{eq:Sscoreex}), $(\hat\beta_2+\hat \beta_3 x_{ij})$ stands for the treatment effect, which depicts the magnitude of the interaction between the variable $X_j$ and the treatment $A$, and $\left[\mathbf{1}(\hat\beta_2+\hat \beta_3 x_{ij}\geq 0)-\hat a\right]$ stands for the proportion of patients whose optimal treatment changes given knowledge of the variable $X_j$. Therefore both these two factors are reflected in S-score, which makes S-score a good quantity that characterizes the degree of qualitative interactions.

\section{Sequential Advantage Selection}

{ S-score characterizes qualitative interaction of each individual covariate, which helps to find prescriptive variables for deriving optimal treatment regimes. However, there are some limitations with the S-score method. First, when the number of covariates is very large, the S-score method tends to select many variables with no qualitative interactions with the treatments since their S-scores are nonzero due to the correlations among covariates. 
Second, since each variable is evaluated individually with the S-score method, some variables that are jointly crucial for optimal treatment decision making may be neglected by the S-score method, because the S-score method does not take correlations between variables into account. We are thus motivated to consider sequential selection based on modified S-scores, which calculates the conditional marginal S-scores for each covariate sequentially with the variables previously included in the model. The new quantity to evaluate each covariate is called {\it sequential advantage}. }

For convenience, we use $\mathcal{M}=\{j^1,..,j^k\}$ to denote  an arbitrary model with $X_{j^1},...,X_{j^k}$ as the selected covariates. Let $\mathcal{F}=\{1,...,p\}$ denote the full model. $\mathbf{X}_i$ is the covariate for $i$th subject and $\mathbf{X}_{i(\mathcal{M})}=\{X_{ij}:j\in \mathcal{M}\}$  is the covariate for $i$th subject corresponding to model $\mathcal{M}$. The sequential advantage of a variable $X_j$ with $k-1$ variables $X_{j^1},...,X_{j^{k-1}}$ already in the model is defined as:
\begin{align*}
S_j^{(k)}=&\frac{1}{n}\sum^n_{i=1}\left\{\max_a\hat{\mathbb{ E}}(Y|\mathbf{X}_{\mathcal{M}_j^{(k)}}=
\mathbf{x}_{i\mathcal{M}_j^{(k)}},A=a)\right.\\
&\left.-\hat{\mathbb{ E}}(Y|\mathbf{X}_{\mathcal{M}_j^{(k)}}=
\mathbf{x}_{i\mathcal{M}_j^{(k)}},A=a_{opt}^{(k-1)}\left(
\mathbf{x}_{i\mathcal{M}^{(k-1)}}\right))\right\},
\end{align*} 
where $\mathcal{M}^{(k-1)}=\{j^1,...,j^{k-1}\}$, $\mathcal{M}_j^{(k)}=\mathcal{M}^{(k-1)}\cup \{j\}$, and $a_{opt}^{k-1}\left(\mathbf{x}_{i\mathcal{M}^{(k-1)}}\right)$ is the optimal treatment regime based on $k-1$ variables $X_{j^1},...,X_{j^{k-1}}$.
This sequential advantage is similar to that in equation (\ref{Sscore}), but the model of the mean response $\mathbb E(Y)$ is based on the current variable $X_j$ and all the variables that are selected before the $k$th step. { Here $\hat{\mathbb E}(Y|\mathbf{X}_{\mathcal{M}_j^{(k)}}=
\mathbf{x}_{i\mathcal{M}_j^{(k)}},A=a)$ is an estimator of the mean response of $Y$ given all the available covariates at this step $\mathbf{X}_{\mathcal{M}_j^{(k)}}$ and treatment $A$. In practice, a linear model with main effects of $\mathbf{X}_{\mathcal{M}_j^{(k)}}$ and $A$, and interaction effects between $\mathbf{X}_{\mathcal{M}_j^{(k)}}$ and $A$ can be used to model $\hat{\mathbb E}(Y|\mathbf{X}_{\mathcal{M}_j^{(k)}},A)$.} This quantity shows the additional advantage of the variable $X_j$ for improving the optimal treatment decision.
Sequential selection guarantees that no redundant variables are included in the decision-making model and leads to a joint model for obtaining the optimal treatment regime.
In the sequential advantage selection, we provide a stopping rule based on the proportion of the incremental sequential advantage. Below is the algorithm of sequential advantage selection for decision making.

{\it Step 1} ({\bf{Initialization}}). Let $\mathcal{M}^{(0)}=\emptyset$. First compute
\begin{equation} \label{eq:step0_opttrt}
a_{opt}^{(0)}=\argmax_a \hat{\mathbb{ E}}(Y|A=a).
\end{equation}
Let 
\begin{equation} \label{eq:step0_S}
S^{(0)}=\hat{\mathbb{ E}}(Y|A=a_{opt}^{(0)})-\{\pi \hat{\mathbb{ E}}(Y|A=1)+(1-\pi)\hat{\mathbb{ E}}(Y|A=0)\},
\end{equation}
where $\pi$ is the probability of assigning a patient treatment 1 in the study. Here $S^{(0)}$ is a quantity of total gain in outcome by using the optimal treatment $a_{opt}^{(0)}$ rather than randomly assigning treatments to patients. This is a baseline advantage increment, and will be used as a reference in the stopping criteria.

{\it Step 2} ({\bf Sequential Advantage Selection}). In the $k$th step ($k\geq 1$), we have $\mathcal{M}^{(k-1)}$. For every $j\in \mathcal{F}\setminus \mathcal{M}^{(k-1)}$, we consider candidate covariates $\mathcal{M}_j^{(k)}=\mathcal{M}^{(k-1)}\cup \{j\}$ and compute the sequential advantage $V$ corresponding to $j$th covariate in $k$th step.
The $k$th variable to be selected is the one with the largest sequential advantage in this step, that is:
\begin{equation} \label{eq:stepk_selvar}
j^k=\argmax_{j \in  \mathcal{F}\setminus \mathcal{M}^{(k-1)}} \{S_j^{(k)}\}.
\end{equation}
Update $\mathcal{M}^{(k)}=\mathcal{M}^{(k-1)}\cup\{j^k\}$. The optimal treatment regime that based on the first $k$ variables $\mathbf{X}_{\mathcal{M}^{(k)}}$ is also updated accordingly:
\begin{equation} \label{stepk_opttrt}
a_{opt}^{(k)}\left(\mathbf{X}_{\mathcal{S}^{(k)}}\right)
=\argmax_a\hat{\mathbb{ E}}(Y|\mathbf{X}_{\mathcal{S}^{(k)}},A=a).
\end{equation}
Let $S^{(k)}=S_{j^k}^{(k)}$, which is the sequential advantage increment based on the $k$th selected variable.

{\it Step 3} ({\bf Stopping Criteria}). Iterate Step 2 until we get the sequence $\mathcal{S}^{p^*}=\{j^1,...,j^{p^*}\}$, where $p^*$ is the first $l\in\{0,1,..,p\}$ that satisfies $\mbox{prop}_{l}\geq c$ and $\mbox{prop}_{l+1}<c$. { Here $\mbox{prop}_l$ stands for the proportion of the incremental sequential advantage and is defined as:
\begin{equation} \label{eq:stopcriteria}
\mbox{prop}_l=\frac{S^{(l)}}{\sum_{i=0}^l S^{(i)}},l=0,1,...,p,
\end{equation}
with $\mbox{prop}_{p+1} = 0$. }
This quantity characterizes the proportion of the additional advantage of the $l$th selected variable to sum of all the sequential advantages explained by the $l$th selected variable and all previously selected variables.
 Here $c$ is a cut-off point for the proportion of the incremental sequential advantage. We choose $c=0.01$, that is, { if a newly selected variable could only explain less than 1\% of the information explained by currently total selected variables, we will not include this variable and stop.} If $\mbox{prop}_1<c$, then $p^*=0$, which means we will not include any variable in the model. The variables $(j^1,...,j^{p^*})$t are the selected important prescriptive variables to make treatment decisions, and $a_{opt}^{(p^*)}(\mathbf{X}_{\mathcal{M}^{(p^*)}})$ is the optimal treatment regime for the corresponding model.

\section{Simulations}
To test performance of the proposed method on variable selection and the accuracy of estimating the decision rule, we compared sequential advantage selection (SAS) with S-score method in \cite{gunter2011variable} and the method proposed by \citet{lu2013variable} under various settings. Although S-score has a natural ranking of the covariates, it doesn't provide an optimal treatment regime based on the variables selected. For fair comparison, we handle S-score method in two ways: one is to include all variables with non-zero S-scores and to see whether they include the true important variables; the other is to include the same number of variables as that of SAS method based on S-score ranking and to fit a new linear model with these variables to get the optimal treatment regime. The method proposed by \citet{lu2013variable} adopts LASSO type shrinkage penalties on {a least squares loss function, which has a form of A-learning and thus does not rely on the correct specification of the baseline function. As done in \citet{lu2013variable}, we use the sample mean of responses $Y$ as the baseline function in A-learning here}, and the LASSO penalty for variable selection. We refer this method as LASSO in simulation results. The tuning parameter $\lambda$ of the LASSO method is chosen by cross validation. The LASSO method and its cross validation is implemented using R-package {\tt glmnet}.  
In our simulation settings, we considered both randomized studies and observational studies with confounders that are not relevant variables for decision making.

For each simulation data set, we generated $n=200$ observations $(\mathbf{X}_i,A_i,Y_i),$ $i=1,...,n$, where $Y_i$ is the response variable, $A_i$ is the treatment assigned to $i$th patient whose possible value is $\{0,1\}$, and  $\mathbf{X}_{i}$ is a column vector containing $p=1000$ independent variables. Let $\mathbf X_{n \times p}=(\mathbf X_1,\mathbf X_2,...,\mathbf X_n)^T$, and $\tilde{\mathbf{X}}_{n\times (p+1)}=(\mathbf{1}_{n \times 1},\mathbf{X})$. We consider the following three models to generate simulation data, which have linear form of covariates for the interaction with treatment, and different function forms for the baseline:
\begin{itemize}
\item Model I: $Y=1+\gamma_1^T \mathbf{X}+A\boldsymbol{\beta}^T\tilde{\mathbf{X}}+\epsilon$ with $\gamma_1=(1,-1,\mathbf{0}_{p-2})^T.$
\item Model II: $Y=1+0.5(\gamma_1^T\mathbf{X})(\gamma_2^T\mathbf{X})+A\boldsymbol{\beta}^T\tilde{\mathbf{X}}+\epsilon$ with
$\gamma_1=(1,-1,\mathbf{0}_{p-2})^T,\gamma_2=(1,\mathbf{0}_2,-1,\mathbf{0}_5,1,\mathbf{0}_{p-10})^T$.
\item Model III: $Y=1+0.5\mbox{sin}(\pi \gamma_1^T\mathbf{X})+0.25(1+\gamma_2^T\mathbf{X})^2+
A\boldsymbol{\beta}^T\tilde{\mathbf{X}}+\epsilon$ with
$\gamma_1$ and $\gamma_2$ being the same as in Model II.
\end{itemize}
Here covariates $\mathbf{X}_i=(X_{i1},...,X_{ip})^T$ are generated from a multivariate normal distribution. Each entry is normal with mean 0, variance 1, and the correlation between covariates is $\mbox{Corr}(X_{ij},X_{ik})=\rho^{|j-k|}$, for $j\neq k$, $j,k=1,...,p$. { In our simulations, we used linear models with main effects of covariates and treatment as well as interaction effects between covariates and treatment to model the response $Y$ given some covariates and treatment. Hence the fitted models in SAS procedure are eventually correctly specified when the true model is Model I, but the main effect parts of the models are misspecified when the true models are Model II and III.} To examine the performance with different correlations between covariates, $\rho$ is chosen to be 0.2, 0.5 and 0.8 to { represent weak, moderate and strong correlations respectively.} The error term $\epsilon$ is normally distributed with mean zero and variance $0.25$.

\begin{figure}[h!]
\centering
\caption{Illustration of qualitative interactions of the covariates $X_1$, $X_9$ and $X_{10}$ with treatment $A$. These are marginal plots of response $Y$ versus covariates $X_1$, $X_9$ and $X_{10}$ (from top to bottom) in two different treatment groups, treatment 1 and treatment 0 from one simulation based on model I and the first choice of $\boldsymbol\beta$. Correlations between covariates are: $\rho=0.2$ (left panel) and $\rho=0.8$ (right panel). The fitted lines are from simple linear regression based on data from one treatment group. Black triangles and dashed lines are from treatment 1, and red circles and dotted lines are from treatment 0.}
\label{qualitative}
\includegraphics[scale=0.5]{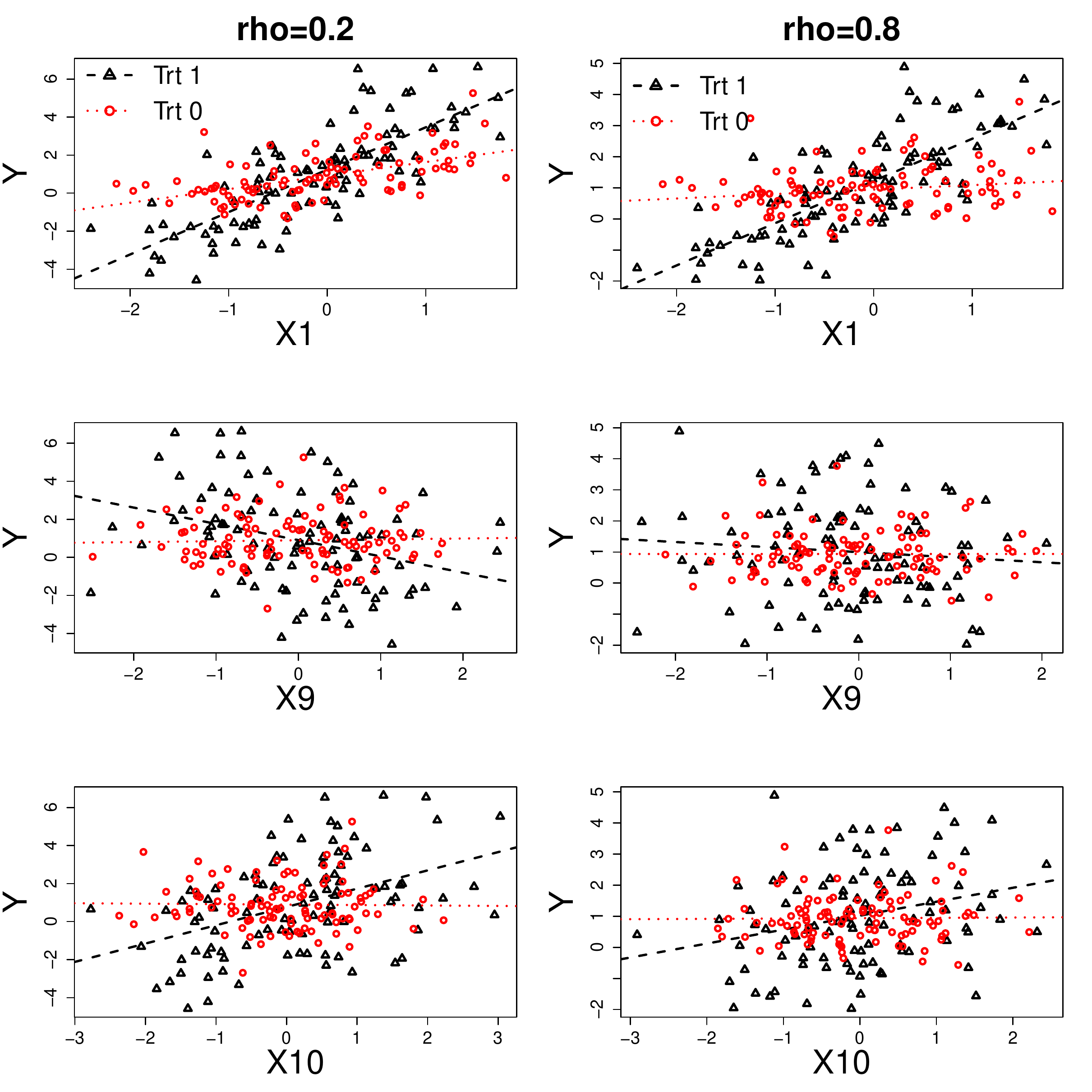}
\end{figure}

Based on these models, the optimal decision will be $I(\boldsymbol\beta^T\tilde {\mathbf X}\geq 0)$, hence important variables for decision making are those whose corresponding coefficient $\beta_j\neq 0$. 
We consider two sets of values for $\boldsymbol\beta$:
\begin{itemize}
\item $\beta_0=0.1,\beta_1=1,\beta_9=-0.9,\beta_{10}=0.8$, other $\beta_j$'s are zero.
\item $\beta_0=0.1,\beta_1=1,\beta_9=-0.9,\beta_{10}=0.8,\beta_{20}=0.8,\beta_{22}=1.5,\beta_{30}=-2,\beta_{35}=2,\beta_{40}=3$, other $\beta_j$'s are zero.
\end{itemize}
In the first scenario, there are 3 important prescriptive variables: $X_1$, $X_9$ and $X_{10}$, { which we refer to as sparse true model.} In the second scenario, there are 8 important prescriptive variables: $X_1$, $X_9$, $X_{10}$, $X_{20}$, $X_{22}$, $X_{30}$, $X_{35}$ and $ X_{40}$, { which we refer to as less sparse true model.} {As the generative models are complex, it becomes rather difficult to evaluate the degree of marginal qualitative interaction of each variable with treatment by the absolute value of $\beta$ due to the correlation between covariates. As an illustration, we show in Figure \ref{qualitative} the marginal interaction plots of variables, $X_1$, $X_9$ and $X_{10}$ with treatment under two choices of correlation between covariates: $\rho=0.2$ and $\rho=0.8$, which are shown in the left and right panel of Figure \ref{qualitative} respectively. These marginal plots are from one simulation under model I with the first choice of $\boldsymbol\beta$. The coefficients for the interaction terms of $X_1$, $X_9$ and $X_{10}$ with treatment $A$ are 1, -0.9 and 0.8. 
Based on Figure \ref{qualitative}, it can be seen that the qualitative interaction effects for one of the variables $X_9$ and $X_{10}$ become weak due to the high correlation between these two variables and the opposite effects of these two variables.} 

In the simulation settings, we consider both scenarios of randomized studies and observational studies. In the randomized study, the treatment is assigned to each patient randomly with $P(A=1)=P(A=0)=1/2$. In the observational study, the treatment is assigned to each patient with $\mbox{logit}\{P(A=1)\}=-0.2+0.8X_1^2+0.8X_{29}^2$, where $\mbox{logit}(u)=\log \{u/(1-u)\}$. 

We implement SAS, S-score and the LASSO methods on each simulated data set with 500 replications. To evaluate the performance of these three methods, we compare them in two aspects. For variable selection, we recorded the average number of variables selected by each method. We also reported true discovery rate (TDR), the proportion correctly identified important variables among all the selected variables, and  true positive (TP), the number of correctly identified important variables. For accuracy of estimated optimal treatment regimes, we evaluated the true {mean response} following the estimated optimal treatment regime $\hat g^{opt}$ of the three methods, as well as the error rate of the estimated optimal treatment regime compared to the true optimal treatment regime. The true value of an estimated treatment regime is estimated by the average outcome of Monte Carlo simulation with 10,000 replicates. Simulation results are shown in Tables \ref{var_i3}, \ref{value_i3}, \ref{var_i8} and \ref{value_i8}, {where Tables \ref{var_i3}, \ref{var_i8} are results for variable selection and Tables \ref{value_i3} , \ref{value_i8} are for results for estimation  of optimal treatment regimes.}

{ We summarize the observations as follows. First, S-score method seems to always select too many variables if all covariates with non-zero S-scores are included. Under weak and moderate correlations, i.e.  $\rho = 0.2$ or $0.5$, the LASSO method tends to select a few more variables than SAS method in models I and II when the true models are sparse, and selected much more variables than SAS method when the true models are less sparse. What's more, standard deviations of the number of variables selected by SAS method are small, which shows stability of SAS method.  As the correlation increases, the average true positives for all three methods gets lower, which is reasonable as there are less information provided by highly correlated covariates. SAS method is capable to include more important variables than the S-score method and LASSO method in all simulation settings. When the true model is sparse,  SAS method can recover almost all the important variables under weak and moderate correlations with small variances of the true positives.  


The true discovery rates depend on both the number of variables selected and the ability to identify important variables. When the true models are sparse, the TDRs of SAS method are not very high especially under moderate to high correlations. However, for the cases where the true models are less sparse, which are shown in Table \ref{var_i8}, the TDRs of SAS method are relatively high because the size of variables selected is small, and are much higher than those of S-score and the LASSO methods.}

{According to Table \ref{value_i3} and Table \ref{value_i8}, SAS method provides a good estimate of optimal treatment regime with values close to the values obtained by implementing the true optimal treatment regime and low error rates, especially in the case of model I. The error rates of S-score method are slightly higher than those of SAS method in most cases. The error rates provided by the LASSO method are approximately 0.4 in most cases. Despite that the LASSO method could select some important variables, the estimated optimal treatment regime did not provide accurate decision rules {\it partly because the LASSO estimates tends to have big bias due to shrinkage}. In Table \ref{value_i8} when the true decision rule is less sparse, the values of estimated optimal treatment regimes of SAS method are the closest to the true optimal values among the three methods in all cases. In Table \ref{value_i3} when the true decision rule is sparse, the values of estimated optimal treatment regimes of the LASSO method are larger than those of SAS method in some cases with Model II and Model III, even if the error rates of the LASSO method are higher. This may be because the values in the cases with only three important variables are not sensitive to treatment regimes, hence a treatment regime that is not close to the optimal treatment regime could still lead to a high value. However, in the case with eight important variables as shown in Table \ref{value_i8}, as the magnitudes of the nonzero coefficients $\boldsymbol\beta$ are larger, treatment regimes that are far from the optimal treatment regime could get values much lower than the optimal value.}

	We also compare solution paths of three methods to evaluate their selection performance. { Here we define the solution path as the trajectory of the number of identified important interaction variables as  the numbers of selected variables increases according to the selection order. For demonstration purpose, we only plot the solution paths for the first 30 selected variables.} SAS method has a natural order of selected variables. For S-score method, we ranked the variables by the descending order of S-scores of these variables. The LASSO method has a solution path of $\boldsymbol\beta$, which can be used to determine the order of variables entering the model. The solution path plots allow us to evaluate the ability of each method to identify important variables given that the same number of variables is selected. These plots are given in Figure \ref{figure_i3r}, \ref{figure_i8r}, \ref{figure_i3o} and \ref{figure_i8o} under different simulation settings.


{ As shown in Figures \ref{figure_i3r}--\ref{figure_i8o}, when the number of variables included in the model for each method is fixed, SAS method includes the largest number of important variables in most cases.
When the data are generated under model I, 
SAS method can include all the important variables in a small number of steps when $\rho$ is not too large. When $\rho$ is large, as the two important variables 9 and 10 are highly correlated, they are likely to be missed by all the three methods, while SAS method still has the highest probability of including these two variables. When the data are generated under model II and model III, all of the methods used the mis-specified models to fit the data and the performance on variable selection is not as good as those in model I. However, SAS method still outperforms other two methods. As the LASSO method is implemented in the framework of A-learning which is robust to baseline function mis-specification, it seems to suggest that SAS method also embraces this robustness.}

{ Overall, SAS method performs well on both the aspect of variable selection and the aspect of estimating optimal treatment regime. SAS method can select most important variables at a moderate size of the set of selected variables. When the model is correctly specified and the correlations between covariates are not too high, the SAS method is able to identify all important variables.  The error rates of the optimal treatment regime based on SAS method are low and the estimated values are close to the truth. While the S-score and LASSO methods have their appeal in some cases, SAS method is also competitive in these cases and is more advantageous considering its ability to identify important variables for decision making.}

\section{Application to STAR*D Study}
We apply the proposed method to data from STAR*D Study, which was conducted to determine the effectiveness of different treatments for patients with major depressive disorder (MDD) who have not been adequately benefiting from initial treatment with an antidepressant.
There are 4041 participants (age 18-75) with nonpsychotic MDD enrolled in this study. Initially these participants were treated with citalopram (CIT) up to 14 weeks. Subsequently, 3 more levels of treatments were provided for participants without a satisfactory response to CIT. At each level, participants were randomly assigned to treatment options acceptable to them. At Level 2, participants are eligible for seven treatment options, which may be conceptualized as two treatment strategies: medication or psychotherapy switch, and medication or psychotherapy augmentation. Available treatments for participants elected to switch are: sertraline (SER), venlafaxine (VEN), bupropion (BUP) and cognitive therapy (CT); available treatments for patients elected to augment are: augmenting CIT with bupropion (CIT+BUP), buspirone (CIT+BUS) or cognitive therapy (CIT+CT). Participants without a satisfactory response to CT were provided additional medication treatments which is called Level 2A. All participants who did not respond satisfactorily at Level 2 or 2A were eligible for Level 3, where possible treatments are medication switch to mirtazapine (MIRT) or nortriptyline (NTP), and medication augmentation with either lithium (Li) or thyroid hormone (THY). Participants without satisfactory response to Level 3 were re-randomized at Level 4 to either tranylcypromine (TCP) or a combination of mirtazapine and venlafaxine (MIRT+VEN). Participants who respond satisfactorily were followed up to 1 year. See \citet{fava2003background} and \citet{rush2004sequenced} for more detailed description of this STAR*D design.

For illustration, we focused on comparing treatment BUP and SER within subset of participants who agreed to be randomized within medication switch at Level 2. Our goal is to identify relevant prescriptive predictors and estimate the optimal treatment decision within this sub group. 
Possible relevant covariates include participant features such as age, gender, socioeconomic status, and ethnicity, illness features such as medication history, family history of mood disorders, and care features such as clinician type. Intermediate medical conditions from Level 1 such as degree of symptom improvement and side effect burden are also considered. 
Table \ref{covariate} lists all of the 305 covariates considered to make treatment decision. Symptomatic status was measured by the 16-item Quick Inventory of Depressive Symptomatology - Clinician-Rated (QIDS-$\mbox{C}_{16}$). Because low QIDS-$\mbox{C}_{16}$ stands for remission, we use negative QIDS-$\mbox{C}_{16}$ as our final outcome. There are 319 participants who had complete records of covariates and final outcomes within medication switch substrategy group that had been randomly assigned to treatment BUP or SER. Among these participants 153 were treated with BUP and 166 were treated with SER.

We applied the three methods, SAS, S-score and the LASSO method considered in simulations to this real data set. The variable selection results are as follows. The LASSO method doesn't select any variable for treatment decision. S-score method had 219 variables with non-zero S-scores. SAS method selected 33 variables with cut-off point in the stopping criteria as 0.01. The reason that the LASSO method didn't include any variables is probably that these covariates could not provide a good prediction on the response. It is reasonable as most of the covariates are categorical, which doesn't include much information about the response. The S-score method selects a lot of covariates, which indicates that there are potentially some prescriptive variables among these covariates. SAS method includes a moderate size of covariates as the important variables, which provides a guidance to clinicians on the covariates that have an effect on treatment decision making. The optimal treatment regime based on SAS method assign 155 patients to treatment BUP, and the rest 164 patients to treatment SER. Among these 33 variables, 10 of them are related to the answers in psychiatric diagnostic screening questionnaire, 5 of them are about rating of depression, 2 of them are QIDS-C and QIDS-SR score changing rates at Level 1, 7 of them are related to patient rated inventory of side effects. The rest 9 variables are demographic factors of the patients. Both covariates at baseline and at Level 1 are selected. For selected covariates at baseline, there are covariates related to both the patient's features and illness features; For selected covariates at Level 1, which are intermediate covariates, there are covariates related to the depression changing rate after the first treatment is given, as well as the side effects of the treatment at Level 1. Hence the optimal treatment decision at Level 2 depends on a comprehensive examination of patient's situation at both baseline and Level 1.

To further examine the variables selected by SAS method, we evaluate the values obtained following the estimated optimal treatment regime based on these selected variables. The average value of a treatment regime is estimated by inverse probability weighted estimator used in \cite{zhang2012robust}, defined as $\mbox{IPW}=\frac{1}{n}\sum_{i=1}^n\frac{Y_iI(A_i=g(\mathbf X_i))}{\pi(A_i)}$. Here $g(\mathbf X_i)$ is the estimated treatment regime, and $\pi(A_i)$ is the probability of receiving treatment $A_i$. As the LASSO method didn't include any variables, we want to compare the estimated value for the estimated optimal treatment regime of SAS method to the values  we get when all subjects are treated with the same treatment, that is, the treatment regime based on no covariate. The estimated value for SAS method is -7.84, and the estimated values with treatment BUP and SER are -10.74 and -10.52, respectively. To examine whether the difference between the value based on SAS method and the value with one treatment only is significantly different from 0, we used 1000 bootstrap sample to obtain the confidence interval for the differences of values. The 95\% bootstrap confidence interval for the value difference between SAS method and BUP treatment is $(0.74,5.15)$, and the 95\% bootstrap confidence interval for the value difference between SAS method and SER treatment is $(0.37,4.82)$. As both confidence intervals don't include zero, it suggested that the value using optimal treatment regime based on SAS method is better than the values based on one treatment only.

\section{Discussion}
In this article, we proposed a variable selection method for optimal treatment decision making, which targets prescriptive variables that are important for decision making and selects variables in a sequential procedure such that the variables selected contain no redundant information.  We provided a stopping criteria such that our final decision-making model can provide enough information with a small number of variables. Our method can be applied to data with continuous outcomes and binary treatment options from a clinical trial or an observational study. Simulation studies demonstrated that our method could identify most important variables under various settings, and provide a good estimated optimal treatment regime that has a small error rate and a high average outcome value. 

{ \cite{gunter2011variable} proposed a hybrid algorithm that combines the LASSO selection and S-score ranking. Specifically, it first uses the LASSO method to select both important  predictive and prescriptive variables, and then uses the S-score method to rank the selected variables for making treatment decision rules. Similarly, SAS method can also be combined with the LASSO method to provide a comprehensive algorithm for variable selection.}  What's more, in the stopping criteria we proposed, we can tune the cut-off point $c$ using data to achieve the goal of maximizing the expected outcome of the treatment regime. 

The proposed method is applicable to the case with more than two treatment options. It may be extended to the situation where the outcome is categorical or censored by appropriately modeling the expected outcome given the covariates and the treatment. Moreover, since dynamic treatment with multiple decision points is common for chronic disease, it is desirable to extend our method to the dynamic treatment regime with multiple stages.


\bibliographystyle{imsart-nameyear}
\bibliography{literature}

\newpage

\newpage

\begin{table}[h!]
\centering
\caption{Results on variable selection of sequential advantage selection (SAS), S-score and LASSO methods for the scenarios with the first choice of $\boldsymbol\beta$ (three important prescriptive variables: $X_1$, $X_9$ and $X_{10}$).
Size is the number of covariates selected. For S-score method, Size $S\neq 0$ is the number of covariates with nonzero S-scores. TDR stands for true discovery rate, which is the proportion correctly identified important variables among all the selected variables. TP stands for true positive, which is the number of correctly identified important variables among $X_1$, $X_9$ and $X_{10}$. The quantity in parenthesis is standard deviation of corresponding value.}
\label{var_i3}
\begin{tabular}{lrrrrrrrrr}
\hline
\multirow{2}{*}{Setting} & \multicolumn{3}{c}{SAS} & \multicolumn{3}{c}{S-score}&\multicolumn{3}{c}{LASSO}\\
&Size  & TDR & TP & Size  $S\neq 0$ &TDR & TP & Size & TDR & TP\\
\hline
Randomized Study &&&&&&&&&\\
Model I, $\rho=0.2$ & 7.67 & 0.42 &2.98& 765.28  & 0.23 &1.66& 17.40 & 0.211&1.60\\
& (2.71) & (0.09) & (0.23) & (163.13) & (0.11) & (0.68) & (23.29) & (0.29) & (1.05) \\
Model I, $\rho=0.5$ & 9.55 & 0.34 &2.92 &757.18&0.16 &1.35& 15.04 & 0.24&1.37\\
& (2.91) & (0.11) & (0.38) & (165.78) & (0.08) & (0.51) & (18.61) & (0.31) & (0.79) \\
Model I, $\rho=0.8$&14.05& 0.17 &2.05&738.44&0.08&1.09& 11.44 & 0.28&1.10\\
& (3.68) & (0.11) & (1.00) & (173.40) & (0.03) & (0.28) & (14.29) & (0.30) & (0.44) \\
Model II, $\rho=0.2$ &23.08&0.12&2.58&762.73&0.11&2.43& 23.42 & 0.22&2.32\\
& (3.87) & (0.04) & (0.76) & (169.38) & (0.04) & (0.62) & (25.22) & (0.23) & (0.86) \\
Model II, $\rho=0.5$ & 22.16 & 0.12&2.41&747.08&0.09 & 1.88& 20.54 & 0.25&1.91\\
& (4.32) & (0.06) & (0.89) & (176.22) & (0.04) & (0.57) & (23.56) & (0.28) & (0.84) \\
Model II, $\rho=0.8$ &17.44 & 0.12&1.91&721.28&0.07 &1.19& 12.93 & 0.27&1.29\\
& (4.56) & (0.08) & (0.98) & (188.22) & (0.04) & (0.40) & (15.93) & (0.27) & (0.58) \\
Model III, $\rho=0.2$ & 22.58 & 0.11&2.35&765.27&0.09&1.98& 15.84 & 0.20&1.48\\
& (4.32) & (0.06) & (1.04) & (165.52) & (0.04) & (0.71) & (21.65) & (0.28) & (1.02) \\
Model III, $\rho=0.5$&23.27 & 0.09&2.04&758.89&0.07&1.46& 13.36 & 0.20& 1.10\\
& (4.04) & (0.06) & (1.02) & (169.00) & (0.03) & (0.57) & (22.14) & (0.29) & (0.83) \\
Model III, $\rho=0.8$ &22.75 & 0.08&1.66&749.84&0.05&1.17& 11.65 & 0.22& 0.98\\
& (4.27) & (0.05) & (0.88) & (170.04) & (0.02) & (0.40) & (15.26) & (0.28) & (0.50) \\
\hline
Observational Study&&&&&&&&&\\
Model I, $\rho=0.2$ & 8.44 & 0.38 &3.00& 760.58  & 0.22&1.73& 34.43 & 0.14&2.94 \\
& (2.13) & (0.09) & (0) & (158.94) & (0.11) & (0.71) & (24.72) & (0.13) & (0.25) \\
 Model I, $\rho=0.5$ &9.51 & 0.35 &2.99& 760.74 & 0.15 &1.36& 36.43 & 0.16&2.71\\
 & (2.89) & (0.11) & (0.15) & (164.11) & (0.08) & (0.57) & (27.92) & (0.20) & (0.58) \\
Model I, $\rho=0.8$ & 12.46 & 0.21 &2.28& 746.41 & 0.10&1.11& 17.63 & 0.28 & 1.62\\
& (3.67) & (0.12) & (0.96) & (167.70) & (0.04) & (0.31) & (21.87) & (0.31) & (0.64) \\
Model II, $\rho=0.2$ & 19.58 & 0.15 &2.84& 527.97  & 0.13&2.42& 28.30 & 0.18&2.96\\
& (4.07) & (0.05) & (0.50) & (207.27) & (0.05) & (0.64) & (22.26) & (0.15) & (0.21) \\
Model II, $\rho=0.5$ &  18.91 & 0.15 &2.67& 609.91  & 0.11 &1.88& 31.67 & 0.16&2.82\\
& (4.00) & (0.06) & (0.72) & (211.18) & (0.04) & (0.57) & (24.06) & (0.17) & (0.45) \\
Model II, $\rho=0.8$& 16.62 & 0.14 &2.08& 672.87  & 0.08  &1.25& 21.04 & 0.25&1.82\\
& (4.21) & (0.08) & (1.00) & (194.65) & (0.04) & (0.44) & (22.84) & (0.28) & (0.78) \\
Model III, $\rho=0.2$ & 19.80 & 0.16 &2.93& 680.88 & 0.10&1.91& 28.68 & 0.18 &2.91\\
& (3.93) & (0.04) & (0.34) & (201.15) & (0.05) & (0.73) & (23.58) & (0.14) & (0.33) \\
Model III, $\rho=0.5$& 20.55 & 0.14 &2.65& 704.98  & 0.08&1.44& 27.82 & 0.19& 2.52\\
& (4.19) & (0.06) & (0.74) & (199.21) & (0.04) & (0.63) & (23.55) & (0.19) & (0.64) \\
Model III, $\rho=0.8$ & 19.66 & 0.11 &2.12& 712.65 & 0.06& 1.13& 16.40 & 0.264&1.64\\
& (4.14) & (0.06) & (0.96) & (191.05) & (0.03) & (0.45) & (19.08) & (0.27) & (0.59) \\
\hline
\end{tabular}
\end{table}

\begin{table}[h!]
\centering
\caption{Estimated optimal values and error rates of sequential advantage selection (SAS), S-score and LASSO methods for the scenarios with the first choice of $\boldsymbol\beta$ (three important prescriptive variables: $X_1$, $X_9$ and $X_{10}$).
True value $Q(g^{opt})$ is the mean response following the true optimal treatment regime $g^{opt}$, while $Q(\hat g^{opt})$ is the mean response following the estimated optimal treatment regime $\hat g^{opt}$ of the three methods. The mean response of each optimal treatment regime is obtained by the average outcome of Monte Carlo simulation using the true model with 10,000 replicates. Err.rate stands for error rate of the estimated optimal treatment regime, which is the proportion of individuals with misspecified optimal treatment following the estimated optimal treatment regime. The quantity in parenthesis is standard deviation of corresponding value.}
\label{value_i3}
\begin{tabular}{lccccccc}
\hline
\multirow{2}{*}{Setting}  &True Value&\multicolumn{2}{c}{SAS} & \multicolumn{2}{c}{S-score} & \multicolumn{2}{c}{LASSO} \\
& $Q(g^{opt})$ & $Q(\hat g^{opt})$ &Err.rate & $Q(\hat g^{opt})$ & Err.rate & $Q(\hat g^{opt})$ & Err.rate\\
\hline
Randomized Study &&&&&&&\\
Model I, $\rho=0.2$ &1.66 
&1.64(0.05)&0.072&1.41(0.13)
&0.275& 1.32(0.19) & 0.436\\
Model I, $\rho=0.5$ &1.60 
&1.56(0.06)&0.103&1.38(0.08)&
0.270 & 1.35(0.15) & 0.449\\
Model I, $\rho=0.8$ &1.51 
&1.41(0.06)&0.182&1.40(0.03)&
0.199& 1.40(0.06) & 0.473\\
Model II, $\rho=0.2$ &2.08 
&1.78(0.10)&0.299&1.79(0.11)&
0.283& 1.45(0.14) & 0.460\\
Model II, $\rho=0.5$ &1.91 
&1.66(0.09)&0.285&1.62(0.08)&
0.289& 1.73(0.10) & 0.466\\
Model II, $\rho=0.8$ &1.65 
&1.49(0.06)&0.243&1.50(0.04)&
0.233& 1.57(0.03) & 0.484\\
Model III, $\rho=0.2$ &2.66 
&2.33(0.13)&0.312&2.30(0.13)&
0.325 & 2.30(0.19) & 0.432\\
Model III, $\rho=0.5$ &2.54 
&2.22(0.10)&0.327&2.18(0.08)&
0.346 & 2.22(0.17) & 0.442\\
Model III, $\rho=0.8$ &2.20 
&1.95(0.06)&0.314&1.98(0.05)&
0.296& 2.03(0.13) & 0.459\\
\hline
Observational Study &&&&&&&\\
Model I, $\rho=0.2$ & 1.66 
&  1.63(0.02) & 0.100 & 1.43(0.12) & 0.257 & 1.56(0.03) & 0.334\\
Model I, $\rho=0.5$ & 1.60 
&  1.55(0.04) & 0.125 & 1.43(0.07) & 0.236 & 1.51(0.03) & 0.319\\
Model I, $\rho=0.8$& 1.51 
& 1.41(0.06) & 0.190 & 1.41(0.02) & 0.196 & 1.45(0.01) & 0.311\\
Model II, $\rho=0.2$ & 2.08 
&  1.82(0.09) & 0.286 & 1.83(0.11) & 0.256 & 2.01(0.04) & 0.314\\
Model II, $\rho=0.5$ & 1.91 
& 1.68(0.08) & 0.278 & 1.66(0.08) & 0.268 & 1.83(0.03) & 0.310 \\
Model II, $\rho=0.8$ & 1.65 
& 1.49(0.06) & 0.251 & 1.51(0.03) & 0.228 & 1.59(0.01) & 0.309\\
Model III, $\rho=0.2$ & 2.66 
&  2.40(0.08) & 0.289 & 2.35(0.13) & 0.289& 2.57(0.04) & 0.320\\
Model III, $\rho=0.5$ & 2.54 
& 2.27(0.09) & 0.303 & 2.23(0.09) & 0.310 & 2.44(0.03) & 0.318\\
Model III, $\rho=0.8$ & 2.20 
& 1.98(0.07) & 0.301 & 2.01(0.06) & 0.273& 2.13(0.01) & 0.312\\
\hline
\end{tabular}
\end{table}

\newpage

\begin{table}[h!]
\centering
\caption{Results on variable selection of sequential advantage selection (SAS), S-score and LASSO methods for the scenarios with the second choice of $\boldsymbol\beta$ (eight important prescriptive variables: $X_1$, $X_9$, $X_{10}$, $X_{20}$, $X_{22}$, $X_{30}$, $X_{35}$ and $X_{40}$).
Size is the number of covariates selected. For S-score method, Size $S\neq 0$ is the number of covariates with nonzero S-scores. TDR stands for true discovery rate, which is the proportion correctly identified important variables among all the selected variables. TP stands for true positive, which is the number of correctly identified important variables among $X_1$, $X_9$, $X_{10}$, $X_{20}$, $X_{22}$, $X_{30}$, $X_{35}$ and $X_{40}$. The quantity in parenthesis is standard deviation of corresponding value.}
\label{var_i8}
\begin{tabular}{lrrrrrrrrr}
\hline
 \multirow{2}{*}{Setting} & \multicolumn{3}{c}{SAS} & \multicolumn{3}{c}{S score} & \multicolumn{3}{c}{LASSO}\\
& Size  & TDR&TP& Size $S\neq 0$ &TDR &TP& Size &TDR&TP\\
\hline
Randomized Study&\\
Model I, $\rho=0.2$ &8.66 & 0.77 &6.36& 775.90 & 0.44&3.63& 32.26 & 0.23&4.99\\
& (2.28) & (0.17) & (1.17) & (166.25) & (0.14) & (0.80) & (22.67) & (0.16) & (1.28) \\
Model I, $\rho=0.5$ & 6.84 & 0.85 &5.63& 776.29  & 0.51&3.38& 32.05 & 0.22&5.03\\
& (1.64) & (0.15) & (0.84) & (164.99) & (0.15) & (0.81) & (20.32) & (0.13) & (1.06) \\
Model I, $\rho=0.8$&5.91 & 0.86 &4.98& 772.77 & 0.29&1.68& 32.80 & 0.19&4.80\\
& (1.13) & (0.15) & (0.85) & (164.44) & (0.10) & (0.61) & (18.94) & (0.11) & (1.07) \\
Model II, $\rho=0.2$ &8.42 & 0.77 &6.22& 773.33 & 0.48&3.81& 34.29 & 0.22&5.32\\
& (2.22) & (0.18) & (1.17) & (160.75) & (0.15) & (0.81) & (22.01) & (0.13) & (1.20) \\
Model II, $\rho=0.5$ & 6.45 & 0.89 &5.61& 776.16 & 0.55&3.42& 34.45 & 0.21&5.30\\
& (1.37) & (0.13) & (0.91) & (163.31) & (0.15) & (0.81) & (22.52) & (0.12) & (1.05) \\
Model II, $\rho=0.8$ &5.65 & 0.89 &5.00& 773.70 & 0.29 & 1.64& 35.02 & 0.18&4.95\\
& (0.91) & (0.13) & (0.79) & (164.49) & (0.11) & (0.59) & (20.24) & (0.10) & (1.01) \\
Model III, $\rho=0.2$ &11.07 & 0.59 &5.94& 775.56 & 0.38&3.82& 31.92 & 0.23&4.91\\
& (3.40) & (0.21) & (1.34) & (164.35) & (0.15) & (0.87) & (23.53) & (0.15) & (1.32) \\
Model III, $\rho=0.5$&9.32 & 0.64 &5.44& 775.36 & 0.43&3.67& 33.28 & 0.22&4.95 \\
& (2.92) & (0.20) & (1.01) & (163.33) & (0.15) & (0.84) & (24.79) & (0.14) & (1.19) \\
Model III, $\rho=0.8$ &6.83 & 0.743 &4.80& 774.23 & 0.28&1.90& 31.35 & 0.20&4.58\\
& (1.83) & (0.20) & (0.91) & (164.92) & (0.10) & (0.69) & (18.45) & (0.13) & (1.16) \\
\hline
Observational Study&\\
Model I, $\rho=0.2$ & 9.11 & 0.76 &6.70& 773.35 & 0.49&4.24& 50.38 & 0.182&7.11\\
& (2.10) & (0.14) & (1.06) & (164.86) & (0.15) & (0.81) & (28.30) & (0.09) & (0.87) \\
Model I, $\rho=0.5$ &7.22 & 0.83 &5.82& 779.48 & 0.55&3.84& 42.70 & 0.20&6.56 \\
& (1.54) & (0.13) & (0.86) & (160.39) & (0.14) & (0.76) & (24.25) & (0.10) & (0.74) \\
Model I, $\rho=0.8$ &  6.21 & 0.80 &4.91& 784.01 & 0.29 &1.76& 40.94 & 0.18&6.02\\
& (1.11) & (0.13) & (0.80) & (157.62) & (0.1) & (0.61) & (19.95) & (0.09) & (0.55) \\
Model II, $\rho=0.2$  & 8.25 & 0.80 &6.34& 705.38 & 0.56&4.34& 48.10 & 0.19& 7.12\\
& (2.18) & (0.18) & (1.18) & (185.85) & (0.16) & (0.84) & (26.87) & (0.10) & (0.91) \\
Model II, $\rho=0.5$ & 6.57 & 0.89 &5.66& 743.78  & 0.60&3.81& 43.33 & 0.20& 6.58\\
& (1.65) & (0.14) & (0.95) & (175.07) & (0.14) & (0.79) & (25.16) & (0.10) & (0.74) \\
Model II, $\rho=0.8$ &  5.61 & 0.91 &5.05& 772.98  & 0.29&1.60& 40.99 & 0.18&6.04\\
& (0.79) & (0.12) & (0.77) & (161.88) & (0.1) & (0.55) & (19.20) & (0.07) & (0.56) \\
Model III, $\rho=0.2$ &10.95 & 0.64 &6.55& 746.16 & 0.43&4.33& 49.26 & 0.19&7.21\\
& (3.11) & (0.18) & (1.19) & (175.46) & (0.14) & (0.86) & (28.36) & (0.09) & (0.84) \\
Model III, $\rho=0.5$&  9.77 & 0.63 &5.69& 756.17  & 0.45&4.06& 44.53 & 0.19&6.68\\
& (3.08) & (0.19) & (1.06) & (167.10) & (0.15) & (0.84) & (25.24) & (0.10) & (0.81) \\
Model III, $\rho=0.8$ & 7.22 & 0.70 &4.79& 770.29 & 0.29&2.06& 41.58 & 0.18&6.15\\
& (2.01) & (0.18) & (0.88) & (161.25) & (0.09) & (0.78) & (19.64) & (0.08) & (0.65) \\
\hline
\end{tabular}
\end{table}

\begin{table}[h!]
\centering
\caption{Estimated optimal values and error rates of sequential advantage selection (SAS), S-score and LASSO methods for the scenarios with the first choice of $\boldsymbol\beta$ (eight important prescriptive variables: $X_1$, $X_9$, $X_{10}$, $X_{20}$, $X_{22}$, $X_{30}$, $X_{35}$ and $X_{40}$).
True value $Q(g^{opt})$ is the mean response following the true optimal treatment regime $g^{opt}$, while $Q(\hat g^{opt})$ is the mean response following the estimated optimal treatment regime $\hat g^{opt}$ of the three methods. The mean response of each optimal treatment regime is obtained by the average outcome of Monte Carlo simulation using the true model with 10,000 replicates. Err.rate stands for error rate of the estimated optimal treatment regime, which is the proportion of individuals with misspecified optimal treatment following the estimated optimal treatment regime. The quantity in parenthesis is standard deviation of corresponding value.}
\label{value_i8}
\begin{tabular}{lccccccc}
\hline
 \multirow{2}{*}{Setting}  &True Value &\multicolumn{2}{c}{SAS} & \multicolumn{2}{c}{S Score}& \multicolumn{2}{c}{LASSO}\\
& $Q(g^{opt})$ & $Q(\hat g^{opt})$ &Err.rate & $Q(\hat g^{opt})$ &Err.rate & $Q(\hat g^{opt})$&Err.rate\\
\hline
Randomized Study &\\
Model I, $\rho=0.2$ &2.94 
&2.86(0.09) & 0.072 & 2.60(0.16) & 0.168& 2.58(0.14) & 0.454\\
Model I, $\rho=0.5$ &2.94 
& 2.87(0.06) & 0.072 & 2.65(0.15) & 0.156& 2.64(0.12) & 0.462\\
Model I, $\rho=0.8$ &2.90 
&2.86(0.04) & 0.058 & 2.49(0.06) & 0.205& 2.64(0.10) & 0.468\\
Model II, $\rho=0.2$ &3.35 
& 3.25(0.07) & 0.089 & 3.08(0.15) & 0.150 & 3.04(0.12) & 0.462\\
Model II, $\rho=0.5$ &3.25 
&3.18(0.04) & 0.075 & 2.98(0.14) & 0.151& 2.98(0.12) & 0.468\\
Model II, $\rho=0.8$ &3.05 
& 3.01(0.03) & 0.056 & 2.63(0.06) & 0.207 & 2.80(0.10) & 0.471\\
Model III, $\rho=0.2$ &3.93 
&3.75(0.15) & 0.117 & 3.58(0.18) & 0.169 & 3.56(0.14) & 0.452\\
Model III, $\rho=0.5$ &3.88 
& 3.73(0.11) & 0.105 & 3.58(0.17) & 0.155 & 3.54(0.14) & 0.457\\
Model III, $\rho=0.8$ &3.60 
&3.52(0.08) & 0.076 & 3.19(0.07) & 0.201& 3.30(0.11) & 0.466\\
\hline
Observational Study &\\
Model I, $\rho=0.2$ & 2.94 
 &  2.89(0.04) & 0.061 & 2.70(0.12) & 0.137 & 2.77(0.06) & 0.313\\
Model I, $\rho=0.5$ & 2.94 
& 2.88(0.03) & 0.066 & 2.73(0.11) & 0.132 & 2.82(0.05) & 0.305\\
Model I, $\rho=0.8$ & 2.90 
&2.86(0.02) & 0.057 & 2.50(0.06) & 0.202 & 2.81(0.05) & 0.300 \\
Model II, $\rho=0.2$ & 3.35 
 &  3.26(0.07) & 0.089 & 3.16(0.09) & 0.129&3.22(0.06) & 0.301\\
Model II, $\rho=0.5$ & 3.25 
& 3.18(0.04) & 0.077 & 3.05(0.11) & 0.131& 3.13(0.05) & 0.299\\
Model II, $\rho=0.8$ & 3.05 
& 3.01(0.02) & 0.056 & 2.63(0.05) & 0.206&2.95(0.05) & 0.299\\
Model III, $\rho=0.2$ & 3.93 
&3.79(0.10) & 0.106 & 3.67(0.13) & 0.144 &3.77(0.07) & 0.306  \\
Model III, $\rho=0.5$ & 3.88 
&3.73(0.09) & 0.107 & 3.65(0.13) & 0.137 &3.74(0.06) & 0.304 \\
Model III, $\rho=0.8$ & 3.60 
&3.53(0.04) & 0.076 & 3.21(0.07) & 0.195  &3.48(0.05) & 0.303\\
\hline
\end{tabular}
\end{table}

\begin{landscape}

\begin{figure}[h!]
\centering
\caption{Solution path of sequential advantage selection (SAS), S-score and LASSO methods. These plots are from {\bf the randomized study} with the first choice of $\boldsymbol\beta$ (three important prescriptive variables: $X_1$, $X_9$ and $X_{10}$), and are given for all combinations of three baseline functions and three choices of correlations of covariates. Black solid line: SAS method; Red dashed line: S-score method; blue dot-dashed line: LASSO method.}
\label{figure_i3r}
\includegraphics[scale=0.6]{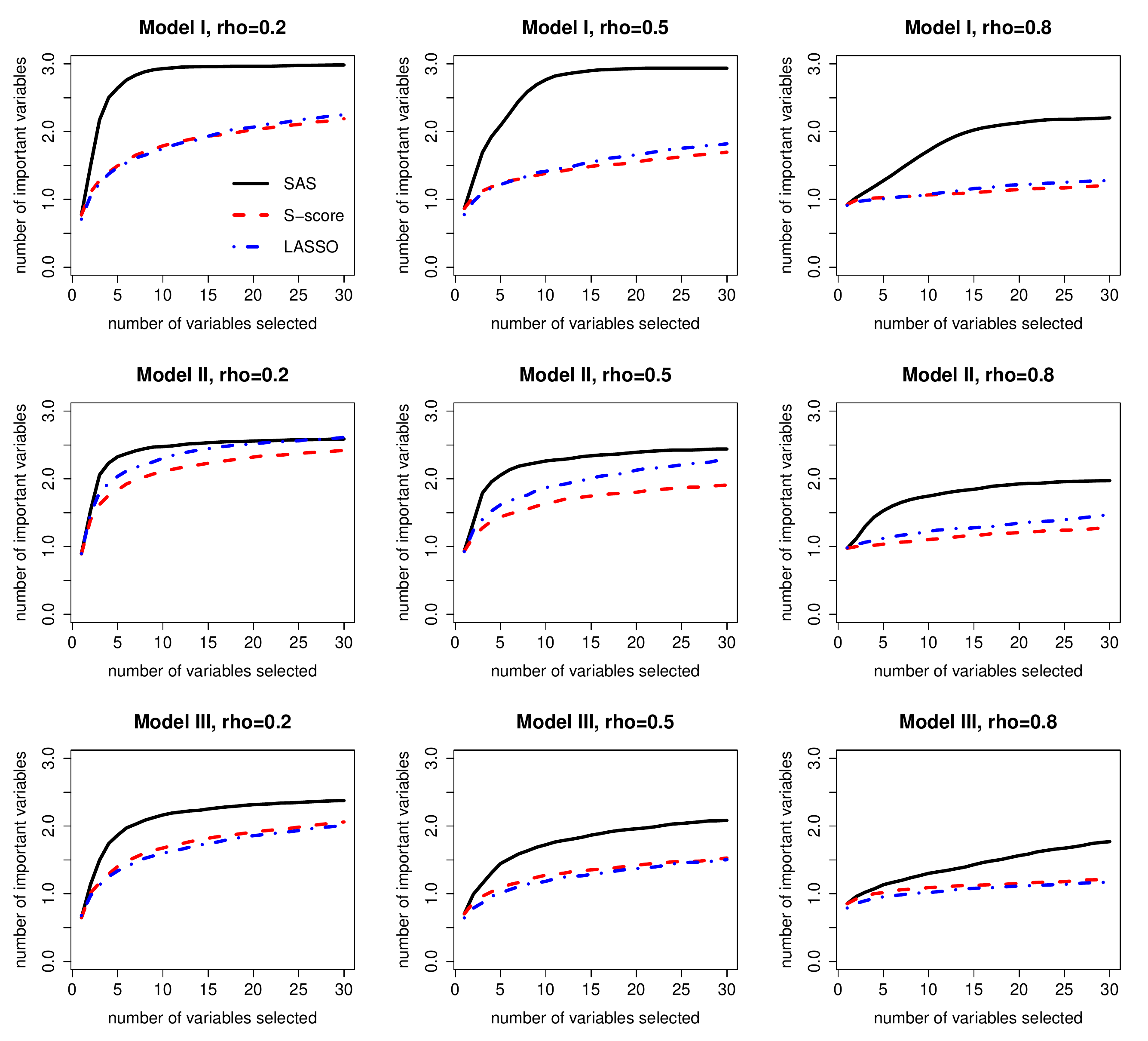}
\end{figure}

\begin{figure}[h!]
\centering
\caption{Solution path of sequential advantage selection (SAS), S-score and LASSO methods. These plots are from {\bf the randomized study} with the second choice of $\boldsymbol\beta$ (eight important prescriptive variables: $X_1$, $X_9$, $X_{10}$, $X_{20}$, $X_{22}$, $X_{30}$, $X_{35}$ and $X_{40}$), and are given for all combinations of three baseline functions and three choices of correlations of covariates. Black solid line: SAS method; Red dashed line: S-score method; blue dot-dashed line: LASSO method.}
\label{figure_i8r}
\includegraphics[scale=0.6]{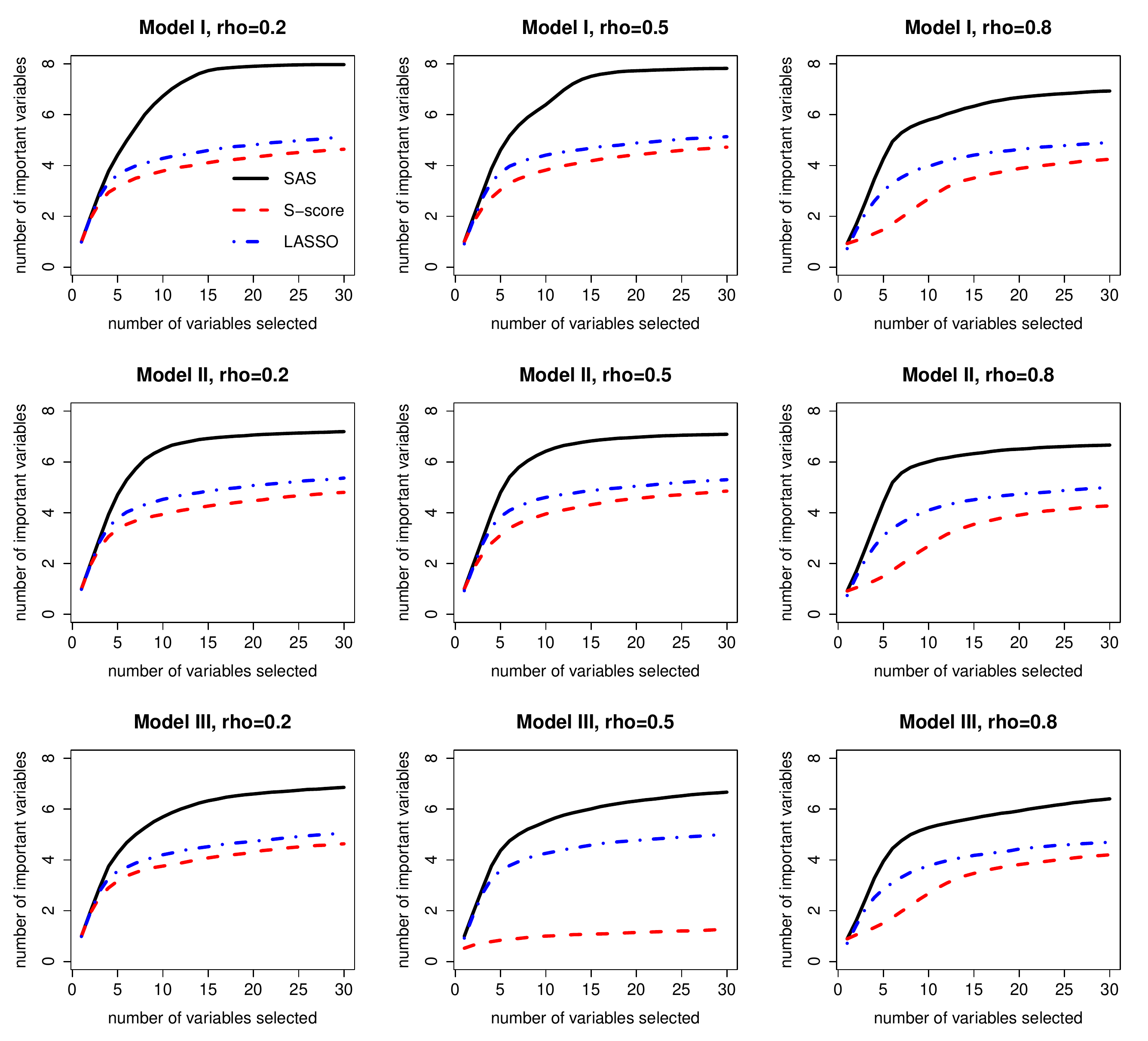}
\end{figure}

\begin{figure}[h!]
\centering
\caption{Solution path of sequential advantage selection (SAS), S-score and LASSO methods. These plots are from {\bf the observational study} with the first choice of $\boldsymbol\beta$ (three important prescriptive variables: $X_1$, $X_9$ and $X_{10}$), and are given for all combinations of three baseline functions and three choices of correlations of covariates. Black solid line: SAS method; Red dashed line: S-score method; blue dot-dashed line: LASSO method.}
\label{figure_i3o}
\includegraphics[scale=0.6]{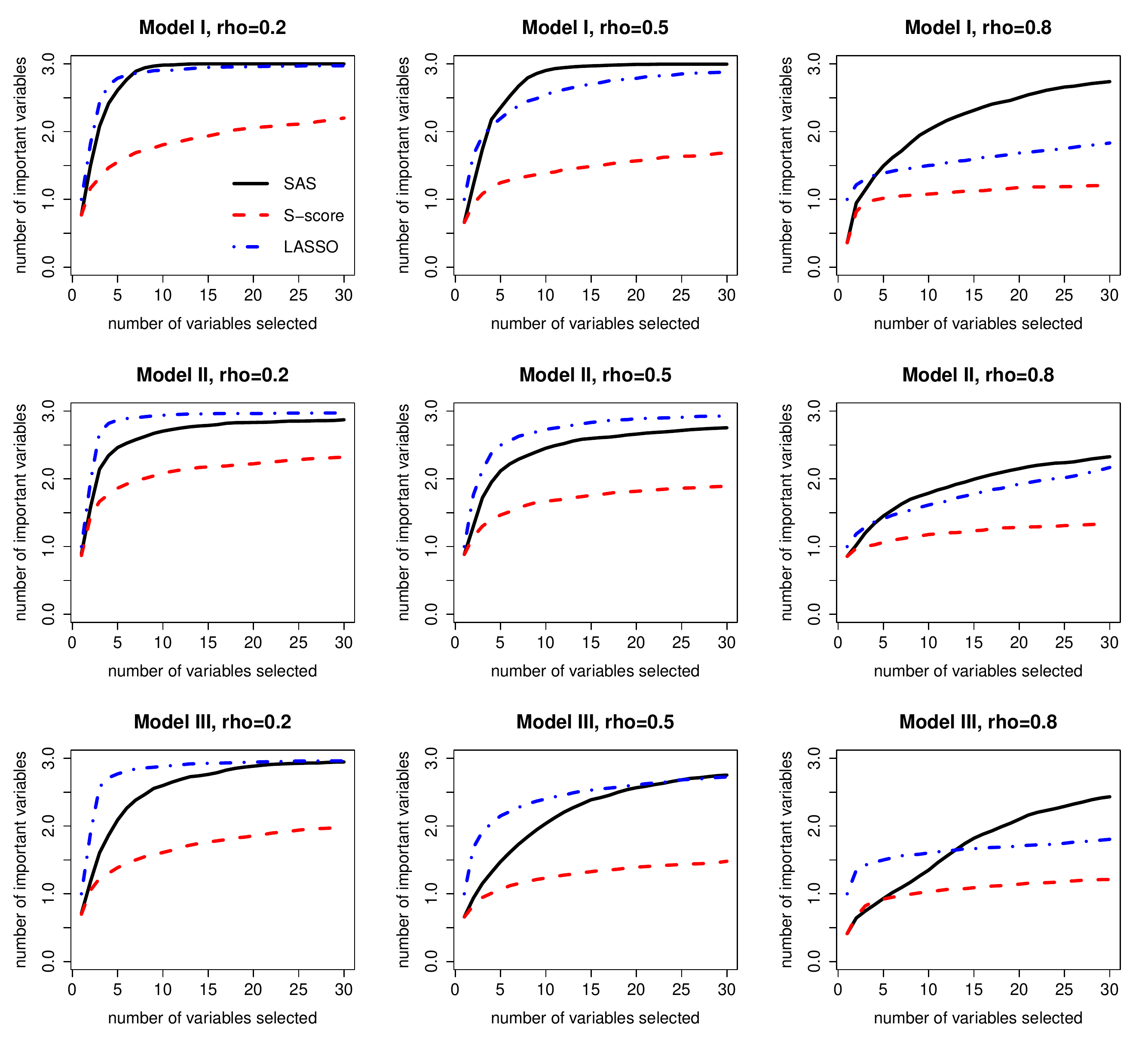}
\end{figure}

\begin{figure}[h!]
\centering
\caption{Solution path of sequential advantage selection (SAS), S-score and LASSO methods. These plots are from {\bf the observational study} with the second choice of $\boldsymbol\beta$ (eight important prescriptive variables: $X_1$, $X_9$, $X_{10}$, $X_{20}$, $X_{22}$, $X_{30}$, $X_{35}$ and $X_{40}$), and are given for all combinations of three baseline functions and three choices of correlations of covariates. Black solid line: SAS method; Red dashed line: S-score method; blue dot-dashed line: LASSO method.}
\label{figure_i8o}
\includegraphics[scale=0.6]{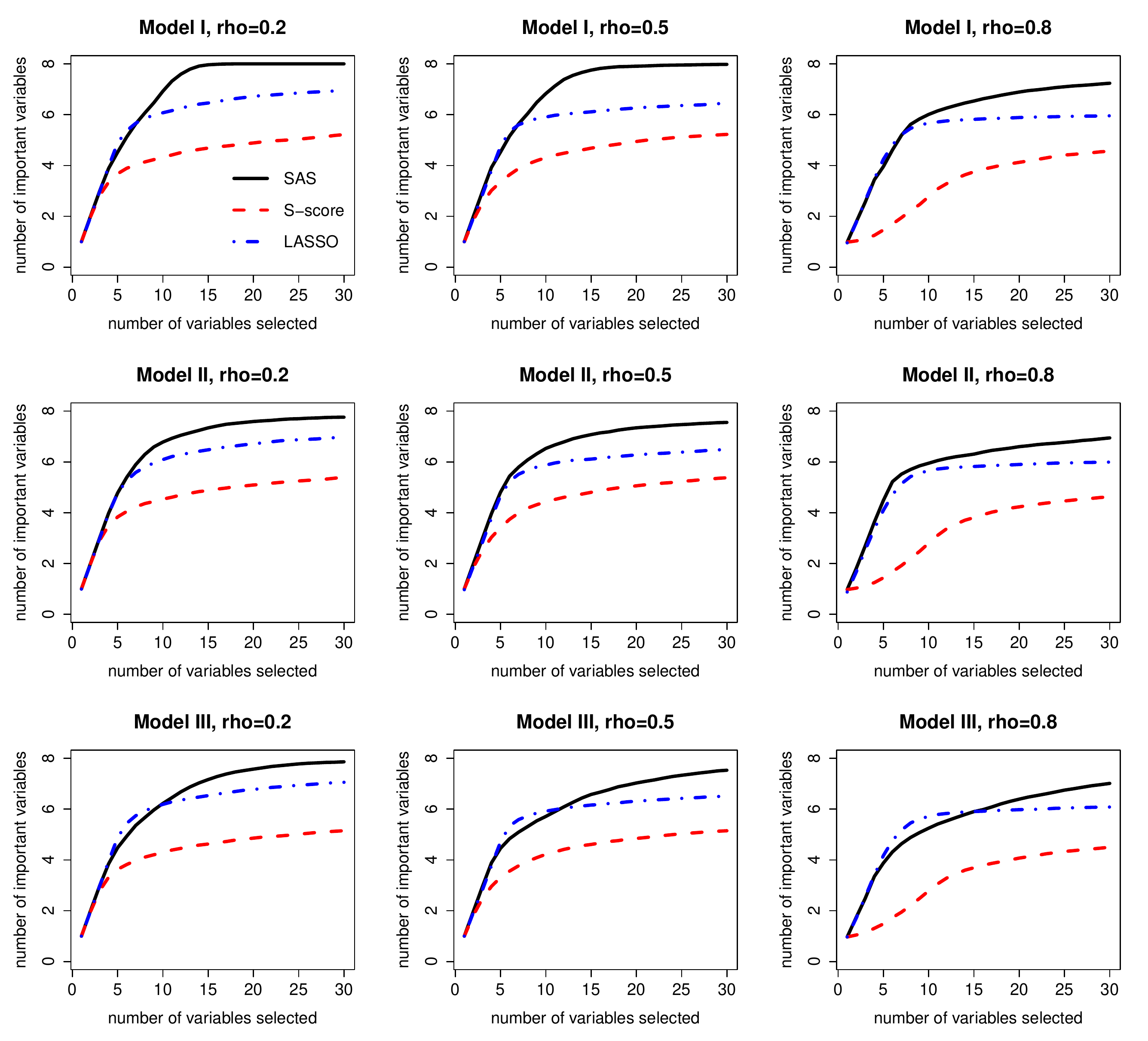}
\end{figure}

\end{landscape}

\begin{table}[t]
\centering
\caption{List of covariates used in application of sequential advantage selection method on the STAR*D study.}
\label{covariate}
\begin{tabular}{ll}
\hline
PARTICIPANT FEATURES & \\
1 Gender & 2-6 Ethnicity\\
7 Economic study consent & 8 Depressed mood\\
9 Diminished interest or pleasure & 10 Weight loss while not dieting\\
11 Insomnia or hypersomnia & 12 Psychomotor agitation or retardation\\
13 Fatigue or loss of energy & 14 Feelings of worthless or guilt\\
15 Diminished ability to concentrate & 16 Recurrent thoughts of death or suicide \\
17 Age & 18 Number of relatives living with patient\\
19 Number of friends living with patient & 20 Total number of persons in household \\
21 Years of schooling completed & 22 Highest degree received\\
23 On medical or psychiatric leave & 24 Medicare \\
25 Medicaid & 26 Private insurance\\
27 Better able make important decisions & 28 Better able to enjoy things  \\
29 Impact of your family and friends & 30-35 Current marital status \\
36-41 Current employment status & 42-44 Currently a student\\
45-46 Currently do volunteer work &\\

ILLNESS FEATURES & \\
47-60 Cumulative Illness Rating Scale &61-78 Hamilton rating scale for depression\\
79-93 Quick Inventory of Depressive Symptomatology & 94-97 Medication history\\
98-236 Psychiatric diagnostic screening questionare & 237 Baseline Axis I psychiatric condition \\
238 Baseline Axis II psychiatric condition & 239 Family hx depression\\
240 Family hx bipolar disorder & 241 Family hx alcohol abuse \\
242 Family hx drug abuse & 243 Family hx suicide\\

INTERMEDIATE MEDICAL CONDITIONS &\\
244-294 Patient rated inventory of side effects & 295 AIDS-C percent improvement \\
296 QIDS-C score change rate & 297 QIDS-SR score change rate \\
298 FISER frequency score change rate & 299 FISER intensity score change rate\\
300 GRSEB score change rate &301 CGII score change rate\\
302 Patient presently a suicide risk & 303 Patient in remission \\
304 Study medical daily dose & \\

CARE FEATURES & \\
305 Type of clinical site &\\
\hline
\end{tabular}
\end{table}

\end{document}